\documentclass{mn2e}
\usepackage{graphicx}
\usepackage{amsbsy}
\usepackage{amsmath}
\usepackage{epsf}
\usepackage{paralist}

\newcommand{\apj}{ApJ}
\newcommand{\apjl}{ApJL}
\newcommand{\mnras}{MNRAS}
\newcommand{\apjs}{ApJS}
\newcommand{\pasp}{PASP}
\newcommand{\aap}{A\&A}

\newcommand{\araa}{ARA\&A}

\newcommand{\etal}{{et al. }}
\newcommand{\beq}{\begin{equation}}
\newcommand{\eeq}{\end{equation}}

\DeclareMathAlphabet{\mathsfsl}{OT1}{cmss}{bx}{sl}
\SetMathAlphabet{\mathsfsl}{bold}{OT1}{cmss}{bx}{sl}

\newcommand*\rfrac[2]{{}^{#1}\!/_{#2}}

\usepackage{epstopdf}
\begin{document}

%===========================================================================
\title[Intrinsic shapes of cores]
{Chemistry as a diagnostic of prestellar core geometry}
%===========================================================================

\author[Tritsis et al.]
  {A.~Tritsis$^{1}$, K.~Tassis$^{1,2}$, K.~Willacy$^{3}$ \\
    $^1$Department of Physics, University of Crete, PO Box 2208, 71003 Heraklion, Greece\\
    $^2$IESL, Foundation for Research and Technology-Hellas, PO Box 1527, 71110 Heraklion, Crete, Greece\\
    $^3$Jet Propulsion Laboratory, California Institute of Technology, Pasadena, CA 91109, USA}

\maketitle 

\begin{abstract}

We present a new method for assessing the intrinsic 3D shape of prestellar cores from molecular column densities. We have employed hydrodynamic simulations of contracting, isothermal cores considering three intrinsic geometries: spherical, cylindrical/filamentary and disk-like. We have coupled our hydrodynamic simulations with non-equilibrium chemistry. We find that\begin{inparaenum}[\itshape a\upshape)] \item when cores are observed very elongated (i.e. for aspect ratios $\le$ 0.15) the intrinsic 3D geometry can be probed by their 2D molecular emission maps, since these exhibit significant qualitative morphological differences between cylindrical and disk-like cores. Specifically, if a disk-like core is observed as a filamentary object in dust emission, then it will be observed as two parallel filaments in $\rm{N_2H^{+}}$; \item for cores with higher aspect ratios (i.e. 0.15 $\sim$ 0.9) we define a metric $\Delta$ that quantifies whether a molecular column density profile is centrally peaked, depressed or flat. We have identified one molecule ($\rm{CN}$) for which $\Delta$ as a function of the aspect ratio probes the 3D geometry of the core; and \item for cores with almost circular projections (i.e. for aspect ratios $\sim$ 1), we have identified three molecules ($\rm{OH}$, $\rm{CO}$ and $\rm{H_2CO}$) that can be used to probe the intrinsic 3D shape by close inspection of their molecular column density radial profiles\end{inparaenum}. We alter the temperature and the cosmic-ray ionization rate and demonstrate that our method is robust against the choice of parameters.   
\end{abstract}


\section{Introduction}\label{intro}
\begin{figure*}
\includegraphics[width=2.2\columnwidth, clip]{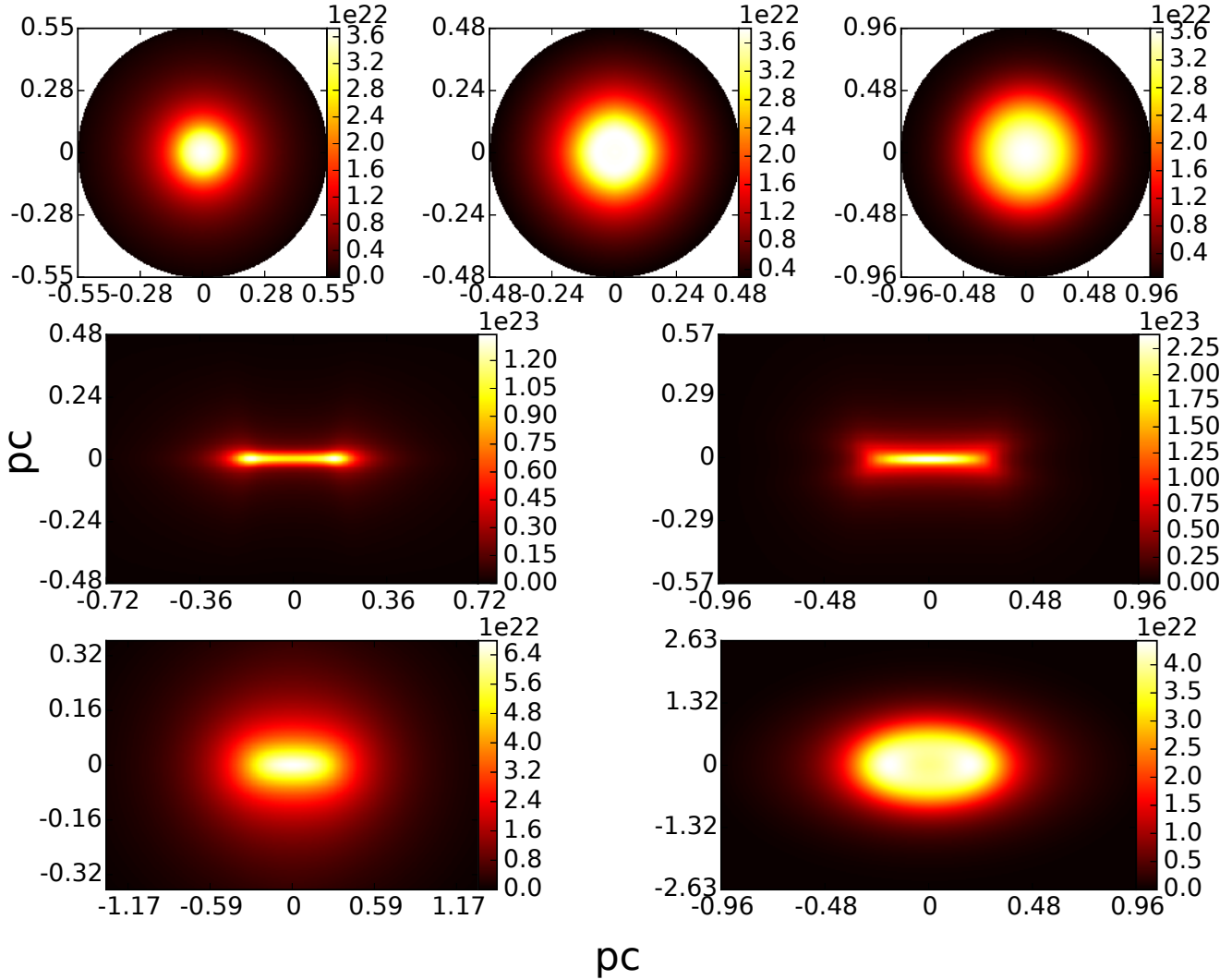}
\caption{Total column density maps from our hydrodynamical simulations for a spherical (upper left), cylindrical (upper middle) and a disk-like (upper right) core as seen face-on. The middle panel depicts the total column density maps for the cylindrical (left) and disk-like (right) cores as seen edge-on. In the lower panels we show total column density maps for the cylindrical (left) and disk-like (right) cores each seen at an angle such that the aspect ratio of their projections is $\sim$ 0.5.
\label{total_cl_high_resolution_final2}}
\end{figure*}

The intrinsic 3D shape of prestellar cores holds important clues about the star formation process since it is determined by the interplay of forces responsible for cloud fragmentation and core formation. Unfortunately, the two dimensional projection of a prestellar core on the plane of the sky probed by dust emission maps can be identical for different intrinsic 3D core shapes (Figure~\ref{total_cl_high_resolution_final2}). Knowledge of the 3D structure of cores combined with observations that probe their kinematics and magnetic field could provide valuable insights as to which is the predominant mechanism that regulates star formation.  

The significance of the problem has lead to numerous statistical studies over the past few years. Myers et al. (1991); Ryden (1996) and Curry(2002) each considered a sample of dense cores and suggested that prestellar cores have a preferentially prolate shape. More recent work (Jones, Basu \& Dubinski 2001; Jones \& Basu 2002; Goodwin; Ward-Thompson \& Whitworth 2002; Tassis 2007; Tassis et al. 2009) has shown that most prestellar cores have an oblate morphology. The theoretical picture is no clearer. In MHD turbulent simulations (Gammie et al. 2003; Li et al. 2004) evidence points towards triaxial cores, with a preference to prolate shapes, whereas simulations where core formation is magnetically driven (Basu \& Ciolek 2004; Ciolek  \&  Basu 2006) indicate that prestellar cores should have oblate shapes. Spherical cores are not favoured in any of the studies mentioned above as they would require all 2D projections to be circular. However, close-to-round objects are observed in nature, albeit seldom. Poidevin et al. (2014) analysed a sample of 27 cores in the Lupus I cloud using {\em Herschel}-SPIRE 350 $\mu$m data. They found that 3 cores had aspect ratios consistent with that of a circular object. Similarly, Tritsis \etal (2015) considered a sample of 27 cores from various clouds and found that 4 had close to circular projections.

For circular objects, as projected on the plane of the sky, dust emission maps alone cannot break the degeneracy between disk-like, cylindrical and spherical cores. Dapp \& Basu (2009) considered analytical column density profiles of prestellar cores and proposed:
\begin{equation}
\label{sphere}
\centering
\Sigma(x)=\frac{\Sigma_c}{\sqrt{1+(x/a)^2}}\times\arctan{\Big(\frac{\sqrt{R^2-x^2}}{\sqrt{x^2+a^2}}\Big)}
\end{equation} 
as a column density profile appropriate for spherical cores. Here, {\em R} is the radius of the sphere, {\em x} is the offset from the centre, $\alpha$ is a parameter proportional to the Jeans length, and $\Sigma_c$ is the central column density. In the same paper they find that the analytical expression for the column density profile for a thin disk viewed with its axis of symmetry parallel to the line of sight (face-on) is:
\begin{equation}
\label{disk}
\centering
\Sigma(x)=\frac{\Sigma_c}{\sqrt{1+(r/a)^2}}
\end{equation} 
where r is the distance from the centre of the disk. For a filamentary cloud, integration of the volume density profile given in Arzoumanian \etal (2011) along the z-axis yields: 
\begin{equation}
\label{cylinder}
\centering
\Sigma_p(r)=\frac{\Sigma_c}{[1+(r/R_{flat})^2]^{p/2}}
\end{equation} 
where $\Sigma_c$ is column density on the axis of symmetry of the cylinder, $R_{flat}$ corresponds to the thermal Jeans length, and the value of p is determined from observations to be $1.5 < \rm{p} < 2.5$. Thus, when the projections of all three geometries considered here are circular, their total column density profiles can in general be characterized by a flat inner part followed by a power-law decrease which steepens at large radii\footnote{Observed column density profiles will deviate from the analytical expressions. A flaring disk seen face-on, will be observed as a ring in a column density map despite the fact that the maximum volume density is at its centre. However, the peak of the ring will be less than a factor of two larger than the column density at the centre of the disk.}. The ambiguity in the total column density profiles between different shapes and projection angles has also been pointed out by past numerical work (Boss \& Hartmann 2001; Ballesteros-Paredes et al. 2003; Nielbock et al. 2012).

Despite the similarity of profiles in dust emission maps, much effort has been made to derive the 3D density distribution of prestellar cores from continuum observations. In a pioneering work, Steinacker et al. (2005) fitted the continuum emission map with Gaussian functions and using an inverse 3D radiative transfer technique they were able to reconstruct the intrinsic structure of the molecular core $\rho$ Oph D. Lomax et al. (2013) fitted dust continuum observations from Ophiuchus, using Bayesian analysis with just one free parameter but with the a priori assumption that the intrinsic shapes were ellipsoids. 

From the chemical point of view, comparison between simulations and molecular observations can provide important clues about the dynamics and the shape of the core. Keto et al. (2015) compared predicted spectra for $\rm{H_2O~(1_{10}-1_{01})}$ and $\rm{C^{18}O~(1-0)}$ with observations from the starless core L1544 for various spherical models. They concluded that the contraction of the core was best approximated by that of a quasi-equilibrium Bonnor-Ebert sphere. Aikawa et al. (2003) were able to reproduce molecular column density profiles for the same core by adopting a Larson-Penston solution for the dynamics.

Molecular observations alone have also been used to estimate the line-of-sight dimensions of structures inside molecular clouds. Li \& Goldsmith(2012) were able to estimate the line-of-sight dimension of the B213 region in Taurus using $\rm{HC_3N}$ (1 - 0) observations. They found that it was comparable to the smaller projected dimension and much smaller than the largest projected dimension, thus suggesting a cylindrical geometry. On the other hand, Storm et al.(2014) derived the line-of-sight dimensions of elongated structures seen in Barnard 1 from the kinematics of the gas. Interestingly enough, they found that the depths into the sky for some of them were comparable to their plane of the sky dimensions, thus suggesting a disk-like geometry.

Here, we present a new recipe designed to break the shape degeneracies caused by projections on the plane of the sky and probe a prestellar core's 3D structure. Our method is based on column densities of commonly observed molecules and can be applied to each core {\em individually}. This paper is organized as follows. In section \S~\ref{simulations} we give an overview of our model. Our chemical network and the parameters altered in our models are described in \S~\ref{chemistry} and \S~\ref{parameters} respectively. We present our results in \S~\ref{results}. We give a summary and discuss our conclusions in \S~\ref{conclusions}.

\section{Numerical methods}

%===========================================================================
\subsection{Models of dynamically evolving cores}\label{simulations}

We have performed hydrodynamic simulations of self-gravitating, isothermal cores in 1D spherical and 2D cylindrical symmetry using the astrophysical code FLASH 4.0.1 (Fryxell et al. 2000; Dubey et al. 2008). We consider 3 intrinsic shapes; spherical, cylindrical/filamentary and disk-like. We solve the equations of hydrodynamics on an adaptive mesh grid with maximum nine levels of refinement (including the zeroth level). The maximum level of refinement yields a resolution of $\sim$10 AU. We use the standard FLASH multipole algorithm to solve Poisson's equation.

For each of our models, the initial density of the core is uniform and equal to $10^3 ~ \rm{cm^{-3}}$. The extent of the computational area for the spherical models is 0.55 pc. In the cylindrical core models the axial and radial dimensions are 0.72 and 0.48 pc respectively. Finally, the size of the simulated region in disk-like core models is 0.96 pc in the radial direction and 0.57 pc in the axial direction. All modelled cores shape have four or more Bonnor-Ebert masses for all temperature values in our parameter study but with their masses being consistent with the core mass function (CMF) (Sadavoy et al. 2010). Thus, all cores are thermally supercritical to collapse. 

Our cylindrical model will eventually fragment in two condensation at its edges with a mean separation of $\sim 0.36$ pc in agreement to observations (Kainulainen et al. 2015). However, for the intended purposes of the current paper, we end our simulations before the fragmentation of the cylinder proceeds at a stage where it cannot longer be considered a continuous structure.

Since our 2D simulations are cylindrically symmetric and in the interest of reducing computational cost we have only simulated one quadrant. At the outer boundary normal velocity components are forced to zero in guard cells so as to have no mass influx. Inner boundaries for these models as well as at both boundaries of our spherical model are reflective. We assume zero initial velocities in all dimensions and allow the cores to collapse under their self-gravity. We terminate each run when the central density reaches $\rm{\sim10^7 cm^{-3}}$.

\begin{table*}
\centering
\caption{\label{species} Chemical species considered}
\scalebox{0.74}{\begin{tabular}{@{}lccccccccccr}
\hline\hline
\multicolumn{11}{c}{Gas phase species} \\
\hline
$\rm{H^{+}}$ & $\rm{H}$ & $\rm{H^{+}_2}$ & $\rm{H^{+}_3}$ & $\rm{He}$ & $\rm{He^{+}}$ & $\rm{C}$ & $\rm{C^{+}}$ & $\rm{CH}$ & $\rm{CH^{+}}$ & $\rm{CH^{+}_2}$ & \\

$\rm{CH_2}$ & $\rm{N}$ & $\rm{N^{+}}$ & $\rm{CH_3}$ & $\rm{NH^{+}}$ & $\rm{CH^{+}_3}$ & $\rm{NH}$ & $\rm{NH^{+}_2}$ & $\rm{O}$ & $\rm{CH_4}$ & $\rm{CH^{+}_4}$ & \\

$\rm{O^{+}}$ & $\rm{NH_2}$ & $\rm{CH^{+}_5}$ & $\rm{OH}$ & $\rm{OH^{+}}$ & $\rm{NH^{+}_3}$ & $\rm{NH_3}$ & $\rm{H_2O}$ & $\rm{NH^{+}_4}$ & $\rm{H_2O^{+}}$ & $\rm{H_3O^{+}}$ & \\

$\rm{C_2}$ & $\rm{C^{+}_2}$ & $\rm{C_2H^{+}}$ & $\rm{C_2H}$ & $\rm{C_2H^{+}_2}$ & $\rm{C_2H_2}$ & $\rm{CN}$ & $\rm{CN^{+}}$ & $\rm{HCN^{+}}$ & $\rm{C_2H^{+}_3}$ & $\rm{HCN}$ & \\

$\rm{HNC}$ & $\rm{Si^{+}}$ & $\rm{C_2H^{+}_4}$ & $\rm{H_2NC^{+}}$ & $\rm{Si}$ & $\rm{N_2}$ & $\rm{CO^{+}}$ & $\rm{HCNH^{+}}$ & $\rm{CO}$ & $\rm{N^{+}_2}$ & $\rm{HCO}$ & \\

$\rm{N_2H^{+}}$ & $\rm{HCO^{+}}$ & $\rm{H_2CO}$ & $\rm{H_2CO^{+}}$ & $\rm{NO}$ & $\rm{NO^{+}}$ & $\rm{H_3CO^{+}}$ & $\rm{CH_3OH}$ & $\rm{O_2}$ & $\rm{O_2^{+}}$ & $\rm{CH_3OH^{+}_2}$ & \\

$\rm{C^{+}_3}$ & $\rm{C_3H^{+}}$ & $\rm{C_2N^{+}}$ & $\rm{CNC^{+}}$ & $\rm{C_3H^{+}_3}$ & $\rm{CH_3CN}$ & $\rm{CH_3CNH^{+}}$ & $\rm{CO_2}$ & $\rm{CO^{+}_2}$ & $\rm{HCO^{+}_2}$ & $\rm{HC_3N}$ & \\

$\rm{HC_3NH^{+}}$ & $\rm{D^{+}}$ & $\rm{D}$ & $\rm{HD^{+}}$ & $\rm{D^{+}_2}$ & $\rm{H_2D^{+}}$ & $\rm{HD^{+}_2}$ & $\rm{D^{+}_3}$ & $\rm{CD}$ & $\rm{CD^{+}}$ & $\rm{CHD^{+}}$ & \\

$\rm{CD^{+}_2}$ & $\rm{CHD}$ & $\rm{CD_2}$ & $\rm{CH_2D}$ & $\rm{CHD_2}$ & $\rm{CD_3}$ & $\rm{ND^{+}}$ & $\rm{CH_2D^{+}}$ & $\rm{CHD^{+}_2}$ & $\rm{CD^{+}_3}$ & $\rm{ND}$ & \\

$\rm{NHD^{+}}$ & $\rm{ND^{+}_2}$ & $\rm{CH_3D}$ & $\rm{CH_2D_2}$ & $\rm{CHD_3}$ & $\rm{CD_4}$ & $\rm{CH_3D^{+}}$ & $\rm{CH_2D^{+}_2}$ & $\rm{CHD^{+}_3}$ & $\rm{CD^{+}_4}$ & $\rm{NHD}$ & \\

$\rm{ND_2}$ & $\rm{CH_4D^{+}_2}$ & $\rm{CH_3D^{+}_2}$ & $\rm{CH_2D^{+}_3}$ & $\rm{CHD^{+}_4}$ & $\rm{CD^{+}_5}$ & $\rm{OD}$ & $\rm{OD^{+}}$ & $\rm{NH_2D^{+}}$ & $\rm{NHD^{+}_2}$ & $\rm{ND^{+}_3}$ & \\

$\rm{NH_2D}$ & $\rm{NHD_2}$ & $\rm{ND_3}$ & $\rm{HDO}$ & $\rm{D_2O}$ & $\rm{NH_3D^{+}}$ & $\rm{NH_2D^{+}_2}$ & $\rm{NHD^{+}_3}$ & $\rm{ND^{+}_4}$ & $\rm{HDO^{+}}$ & $\rm{D_2O^{+}}$ & \\

$\rm{H_2DO^{+}}$ & $\rm{HD_2O^{+}}$ & $\rm{D_3O^{+}}$ & $\rm{C_2D{+}}$ & $\rm{C_2D}$ & $\rm{C_2D^{+}_2}$ & $\rm{C_2HD}$ & $\rm{C_2D_2}$ & $\rm{DCN^{+}}$ & $\rm{C_2H_2D^{+}}$ & $\rm{C_2HD^{+}_2}$ & \\

$\rm{C_2D^{+}_3}$ & $\rm{DCN}$ & $\rm{DNC}$ & $\rm{C_2H_3D^{+}}$ & $\rm{C_2H_2D^{+}_2}$ & $\rm{C_2HD^{+}_3}$ & $\rm{C_2D^{+}_4}$ & $\rm{HDNC^{+}}$ & $\rm{D_2NC^{+}}$ & $\rm{DCNH^{+}}$ & $\rm{HCND^{+}}$ & \\

$\rm{DCND^{+}}$ & $\rm{DCO}$ & $\rm{N_2D^{+}}$ & $\rm{DCO^{+}}$ & $\rm{HDCO}$ & $\rm{D_2CO}$ & $\rm{HDCO^{+}}$ & $\rm{D_2CO^{+}}$ & $\rm{H_2DCO^{+}}$ & $\rm{HD_2CO^{+}}$ & $\rm{D_3CO^{+}}$ & \\

$\rm{CH_2DOH}$ & $\rm{CHD_2OH}$ & $\rm{CD_3OH}$ & $\rm{CH_3OD}$ & $\rm{CH_2DOD}$ & $\rm{CHD_2OD}$ & $\rm{CD_3OD}$ & $\rm{CH_3OHD^{+}}$ & $\rm{CH_3OD^{+}_2}$ & $\rm{CH_2DOH^{+}_2}$ & $\rm{CHD_2OH^{+}_2}$ & \\

$\rm{CD_3OH^{+}_2}$ & $\rm{CH_2DOHD^{+}}$ & $\rm{CHD_2OHD^{+}}$ & $\rm{CD_3OHD^{+}}$ & $\rm{CH_2DOD^{+}_2}$ & $\rm{CHD_2OD^{+}_2}$ & $\rm{CD_3OD^{+}_2}$ & $\rm{C_3D^{+}}$ & $\rm{C_3H_2D^{+}}$ & $\rm{C_3HD^{+}_2}$ & $\rm{C_3D^{+}_3}$ & \\

$\rm{CH_2DCN}$ & $\rm{CHD_2CN}$ & $\rm{CD_3CN}$ & $\rm{CH_3CND^{+}}$ & $\rm{CH_2DCNH^{+}}$ & $\rm{CHD_2CNH^{+}}$ & $\rm{CD_3CNH^{+}}$ & $\rm{CH_2DCND^{+}}$ & $\rm{CHD_2CND^{+}}$ & $\rm{CD_3CND^{+}}$ & $\rm{DCO^{+}_2}$ & \\

$\rm{DC_3N}$ & $\rm{DC_3NH^{+}}$ & $\rm{HC_3ND^{+}}$ & $\rm{DC_3ND^{+}}$ & $\rm{HD}$ & $\rm{CHD_2CNH^{+}}$ & $\rm{CD_3CNH^{+}}$ & $\rm{CH_2DCND^{+}}$ & $\rm{CHD_2CND^{+}}$ & $\rm{CD_3CND^{+}}$ & $\rm{DCO^{+}_2}$ & \\

$\rm{C_3H_2}$ & $\rm{C_3H}$ & $\rm{C_3H^{+}_2}$ & $\rm{C_3HD}$ & $\rm{C_3D}$ & $\rm{C_3HD^{+}}$ & $\rm{C_3D_2}$ & $\rm{C_3D^{+}_2}$ & $\rm{H_2}$ & $\rm{D_2}$ &  & \\
\hline
\multicolumn{11}{c}{Dust grain species} \\
\hline
$\rm{H}$ & $\rm{C}$ & $\rm{CO}$ & $\rm{H_2CO}$ & $\rm{Si}$ & $\rm{C_2}$ & $\rm{O_2}$ & $\rm{CH}$ & $\rm{OH}$ & $\rm{NO}$ & $\rm{CH_2}$ & \\

$\rm{H_2O}$ & $\rm{CO_2}$ & $\rm{CH_3}$ & $\rm{CH_4}$ & $\rm{HNC}$ & $\rm{HCO}$ & $\rm{C_2H_2}$ & $\rm{HC_3N}$ & $\rm{N_2}$ & $\rm{CN}$ & $\rm{NH}$ & \\

$\rm{HCN}$ & $\rm{C_2H}$ & $\rm{NH_3}$ & $\rm{CH_3CN}$ & $\rm{CH_3OH}$ & $\rm{NH_2}$ & $\rm{N}$ & $\rm{O}$ & $\rm{H_2}$ & $\rm{CH_2OH}$ & $\rm{D}$ & \\

$\rm{HDCO}$ & $\rm{D_2CO}$ & $\rm{CD}$ & $\rm{OD}$ & $\rm{CHD}$ & $\rm{CD_2}$ & $\rm{HDO}$ & $\rm{D_2O}$ & $\rm{CH_2D}$ & $\rm{CHD_2}$ & $\rm{CD_3}$ & \\

$\rm{CH_3D}$ & $\rm{CH_2D_2}$ & $\rm{CHD_3}$ & $\rm{CD_4}$ & $\rm{DNC}$ & $\rm{DCO}$ & $\rm{C_2HD}$ & $\rm{C_2D_2}$ & $\rm{DC_3N}$ & $\rm{ND}$ & $\rm{DCN}$ & \\

$\rm{C_2D}$ & $\rm{NH_2D}$ & $\rm{NHD_2}$ & $\rm{ND_3}$ & $\rm{CH_2DCN}$ & $\rm{CHD_2CN}$ & $\rm{CD_3CN}$ & $\rm{CH_2DOH}$ & $\rm{CHD_2OH}$ & $\rm{CD_3OH}$ & $\rm{CH_3OD}$ & \\

$\rm{C_2D}$ & $\rm{NH_2D}$ & $\rm{NHD_2}$ & $\rm{ND_3}$ & $\rm{CH_2DCN}$ & $\rm{CHD_2CN}$ & $\rm{CD_3CN}$ & $\rm{CH_2DOH}$ & $\rm{CHD_2OH}$ & $\rm{CD_3OH}$ & $\rm{CH_3OD}$ & \\	

$\rm{CH_2DOD}$ & $\rm{CHD_2OD}$ & $\rm{CD_3OD}$ & $\rm{NHD}$ & $\rm{ND_2}$ & $\rm{HD}$ & $\rm{D_2}$ & $\rm{CHDOH}$ & $\rm{CD_2OH}$ & $\rm{CH_2OD}$ & $\rm{CHDOD}$ & \\	

$\rm{CD_2OD}$ & $\rm{C_3H_2}$ & $\rm{C_3H}$ & $\rm{C_3HD}$ & $\rm{C_3D_2}$ & $\rm{C_3D}$ &  &  &  &  &  & \\	
\hline\hline
\end{tabular}}
\end{table*}

%============================================================================

\subsection{Chemical Network}\label{chemistry}

We couple the dynamical models with non-equilibrium chemistry and follow the abundances of 214 gas-phase and 82 dust grain species. The evolution of these species is governed by 13967 chemical reactions. The reaction rates of our chemical network are adopted from the fifth release of the UMIST database (McElroy et al. 2013). In Table~\ref{species} we list all the species present in our network. 

The initial elemental abundances, relative to the total density, are $\rm{[He]=2.2\times 10^{-1}}$, $\rm{[Si^{+}]=3.1\times 10^{-8}}$, $\rm{[N]=3.3\times 10^{-5}}$, $\rm{[C^{+}]=1.1\times 10^{-4}}$, and $\rm{[O]=2.7\times 10^{-4}}$. Approximately 98\% of hydrogen is in molecular form with the rest as H atoms.  The total deuterium abundance is $1.87\times 10^{-5}$, with 33\% in molecular form and 66\% in the form of $\rm{HD}$. Thus, the C/O ratio is 0.4, the D/H ratio $\rm{1.6\times 10^{-5}}$, and the mean molecular weight is 2.4. The initial abundances of molecular species are $\rm{[H_2]=7.6\times 10^{-1}}$ and $\rm{[D_2]=6.2\times 10^{-6}}$. Therefore, at the beginning of each simulation, the only molecules present are $\rm{H_2}$, $\rm{D_2}$ and HD.

FLASH is able to monitor multiple fluids. We have appropriately modified the ionization unit which already has multispecies capabilities build in, and treat each molecular species as a different fluid. Thus, for each species a separate advection equation is solved with chemical abundances calculated after advection terms. The abundances of $\rm{H_2}$ and $\rm{D_2}$ are calculated from the conservation of the total hydrogen and deuterium at the end of each timestep. We follow the formulation of Tassis \etal (2012) to model gas-grain interactions.

The abundance profiles from our chemical model were compared with those of previous chemical studies (Tassis et al. 2012) and were found to be in good agreement. The results of these previous chemical runs were in turn extensively compared with observations from a number of dense cores (see Tassis et al. 2012 for a list of the observational studies). The values obtained from these observations are, within uncertainties, consistent with the range of predictions obtained by the chemical models.
%============================================================================
\subsection{Parameter Study}\label{parameters}

We study the sensitivity of chemical abundances on the temperature and the cosmic-ray ionization rate by performing a suite of simulations. The parameters altered in each run are listed in Table~\ref{Models}.

For our reference run (i.e. run 1 in Table~\ref{Models}) we adopt typical $\rm{H_2}$ cloud conditions. The temperature is set at 10 K, and a standard value of $\rm{\zeta=1.3\times 10^{-17}~s^{-1}}$ is used for the cosmic-ray ionization rate. For each shape, we consider models with $\rm{T=7 ~ K}$ and $\rm{T=14 ~ K}$. We change the cosmic-ray ionization rate a factor of four above and bellow the standard value ($\rm{\zeta=5.2\times 10^{-17}~s^{-1}}$ and $\rm{\zeta=3.3\times 10^{-18}~s^{-1}}$ respectively). Thus, we have a total of 15 runs, 5 for each intrinsic geometry. The initial abundances of metals are the same for all runs.
\begin{table}
\centering
\caption{Parameters used in each run.}
\label{Models} 
\begin{tabular} {@{}c c c}
\hline\hline
Runs & Temperature (K) & Ionization Rate ($s^{-1}$) \\
\hline
1 & 10 & $1.3\times 10^{-17}$ \\
2 & 14 & $1.3\times 10^{-17}$ \\
3 & 7 & $1.3\times 10^{-17}$ \\
4 & 10 & $5.2\times 10^{-17}$ \\
5 & 10 & $3.3\times 10^{-18}$ \\
\hline\hline
\end{tabular}
\end{table}

%============================================================================

\section{Results}\label{results}
\begin{figure}
\includegraphics[width=1.0\columnwidth, clip]{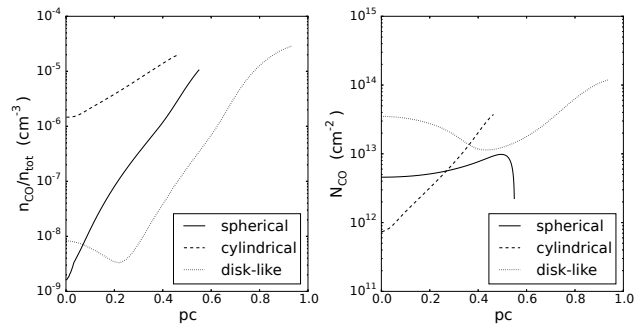}
\caption{Radial profiles of CO abundance (left) and CO column density (right) for the three geometries. The disk-like and cylindrical core models are assumed to be viewed face-on. Abundance profiles for the disk-like and cylindrical cores are taken to extend radially from the maximum density point on the axis of symmetry. The central density for all three geometries is $\rm{10^6 ~ cm^{-3}}$.}
\label{CO_radia_cl_vs_radial_ab4}
\end{figure}

We have produced and examined the column density maps, in various projection angles, of the total density and of all molecules present in our chemical network when the central density for all three geometries is $\rm{10^6 ~ cm^{-3}}$. The total column density maps of our simulated cores when these appear circular (i.e with aspect ratios $\sim$ 1), very elongated (i.e. with aspect ratios $\le$ 0.15), and with an aspect ratio of $\rfrac{1}{2}$ are shown in the upper, middle and lower panels of Figure~\ref{total_cl_high_resolution_final2} respectively. 
\begin{figure*}
\includegraphics[width=2.1\columnwidth, clip]{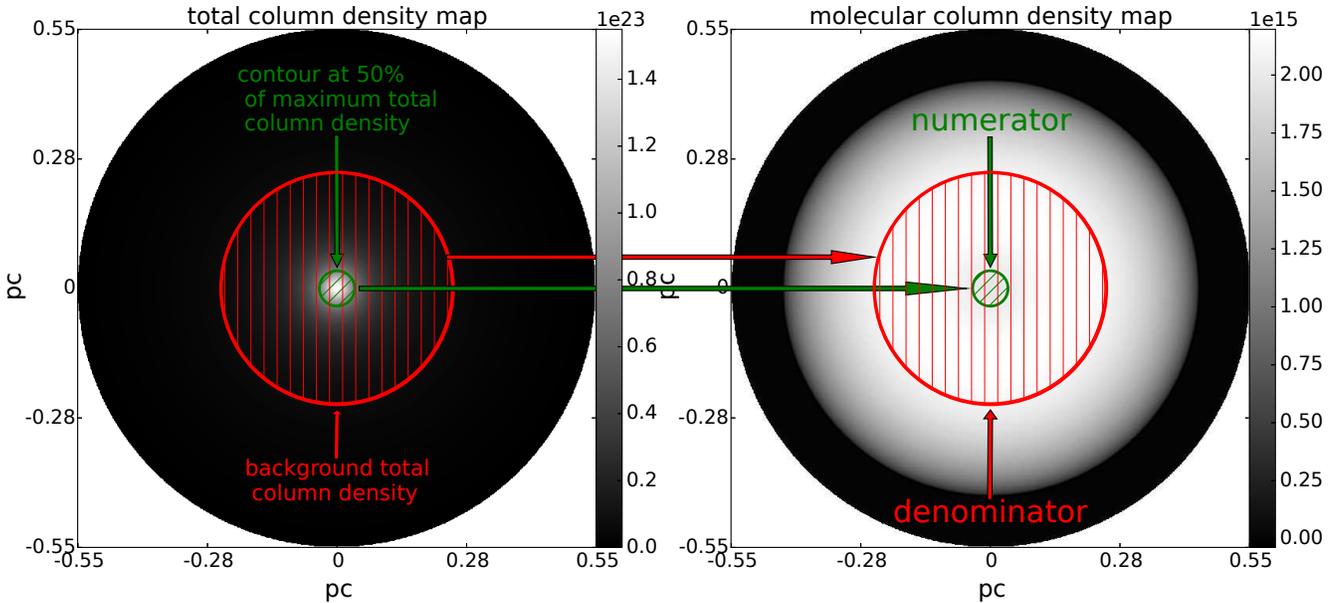}
\caption{Schematic definition of the parameter $\Delta$ for a circular object. In the left panel we plot the total column density of our spherical model. The green circle is a contour that marks 50\% of the maximum of the {\em total column density}. The numerator of $\Delta$ is the mean of the molecular column density inside that contour ({\em green hatched region} in the right panel). The red circle is a contour that marks the region of the core with total column density higher than a background value. The denominator of $\Delta$ is the mean of the molecular column density inside the red circle ({\em red hatched region} in the right panel). The molecular column density map here is that of $\rm{CO}$ for a spherical geometry.} 
\label{Delta_explained3}
\end{figure*}

A two-dimensional projection of a spherical core will always be circular. Hence, candidates for spherical cores are the easiest to identify. Nonetheless, an indistinguishable, circular shape can also result from the projection of a cylindrical or a disk-like cloud if viewed with their axis of symmetry parallel to line of sight (face-on) (see upper panel of Figure~\ref{total_cl_high_resolution_final2}). Therefore, even seeing a core as a close-to-round object in a dust emission map does not necessarily imply it is spherical. In the opposite case, when a disk-like and a cylindrical core are observed with their axis of symmetry perpendicular to the line of sight (edge-on), they would both appear as elongated objects almost identical to one another (see middle panel of Figure~\ref{total_cl_high_resolution_final2}). The degeneracy amongst the 2D projections of a disk-like and a cylindrical cloud remains for intermediate projection angles as well. In this case, both intrinsic 3D shapes would manifest themselves as ellipsoids with their aspect ratios depending on the projection angle. This is shown in the lower panel of Figure~\ref{total_cl_high_resolution_final2} when the cylindrical (left) and the disk-like cores (right) are both viewed such that their projections have aspect ratios $\rfrac{1}{2}$. For the same evolutionary stage, the projection angle required so that our simulated cylindrical and disk-like cores are seen as ellipses with aspect ratio $\rfrac{1}{2}$ would be 85$^{\circ}$ and 65$^{\circ}$ respectively. However, the frequency of the different aspect ratios varies for different intrinsic shapes and has been studied before (Curry 2002; Tassis 2007). Although our results are consistent with these studies, in the edge-on case these objects would be most probably identified as dense "filaments" rather than "cores" with the difference only being semantic.

Physical scales in Figure~\ref{total_cl_high_resolution_final2} are of little importance since in real life, a core of smaller size could be at a smaller distance and thus appear equal in size to a larger core at a larger distance. Depending on the mass and the evolutionary stage, two cores with different intrinsic geometries, located at the exact same distance, could have the same physical scales.

A molecular column density profile is determined by two factors: 
\begin{enumerate}
\item the distribution of the abundance of that molecule, and
\item the path of integration along a line-of-sight.
\end{enumerate}
Clouds and cores with different intrinsic shapes evolve at different rates. Since the dynamical and chemical evolution are coupled (Aikawa \etal 2002; Tassis \etal 2012) the abundance of a molecular species and its distribution at a certain evolutionary stage will differ for discrete geometries. The path of integration along a line of sight is also subject to the core's morphology. We illustrate these two effects in Figure~\ref{CO_radia_cl_vs_radial_ab4} assuming the disk-like and cylindrical cores are seen face-on. In the left panel we plot the abundance of $\rm{CO}$ and for all three intrinsic geometries the profiles are centrally depressed as expected for a depletion-affected molecule. In the right panel, we show the column density of $\rm{CO}$ for these geometries. For the spherical core it is only slightly centrally depressed, for the disk-like core it is slightly centrally peaked with an overall flat behaviour, and only for the cylindrical core does it resemble the true, abundance profile of $\rm{CO}$. 

It is clear that geometrical effects have a strong impact on the appearance of the core when it is observed through the chemical lens. The differences induced in molecular column density profiles from the intrinsic shape of the core can in turn be used to identify its true shape. However, radiative transfer processes also have a strong influence on how a core is observed on the plane of the sky. Such processes are approximated by either the critical density ($\rm{n_{crit}}$) or by the effective density ($\rm{n_{eff}}$). 

\begin{figure}
\includegraphics[width=1.0\columnwidth, clip]{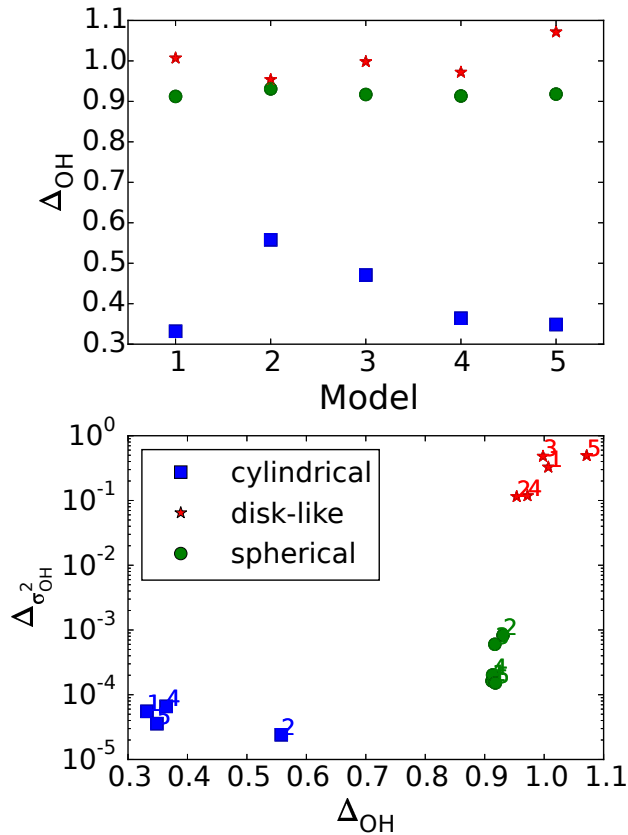}
\caption{Upper panel: the value of the parameter $\Delta$ for $\rm{OH}$ for each run and each geometry separately. Lower panel: The parameter $\Delta_{\sigma^2}$ plotted against the parameter $\Delta$. Symbols represent different intrinsic geometries (red star=disk-like, green circle=sphere, and blue square=cylindrical) and numbers denote runs with different parameters (see Table~\ref{Models}). Different intrinsic shapes occupy different regions of the plot.}
%Model one is our reference model, in two the temperature is $T=14 ~K$, in three $T=7 ~K$, in four the ionization degree is a factor of four above the standard value and in five a factor of four bellow.}
\label{main_combo2}
\end{figure}

The concept of critical density is an overly simplistic approximation since subthermal excitation of molecules is neglected. In addition, calculations of $\rm{n_{crit}}$ usually do not take under consideration optical depth effects. The effective density is defined as the density needed to produce an 1 K molecular line (Evans 1999). This is an easily detectable line. In contrast to $\rm{n_{crit}}$, in $\rm{n_{eff}}$ radiative trapping is accounted for. The concept of $\rm{n_{eff}}$ is not free of caveats either. The approximation brakes down for low molecular column densities and is not appropriate to describe more complex molecules ($\rm{CH_3CHO}$). However, we examine prestellar cores where the molecular column density is by definition high and we do not propose any complex molecule as a geometry tracer.

We produced our molecular column density maps by considering only the regions of the core with density higher than the effective excitation density. We repeated the same analysis by considering the critical density $\rm{n_{crit}}$ instead of $\rm{n_{eff}}$. We present only the molecules for which the results with $\rm{n_{crit}}$ and with $\rm{n_{eff}}$ converge to each other. 
In this manner, we ensure that radiative transfer effects are negligible compared to the chemical effects. Critical and effective densities are adopted from Shirley (2015). Driven from observations (Marchwinski et al. 2012) and the discussion in Shirley (2015), we adopt a value of $\rm{10^2~cm^{-3}}$ for the effective density of $\rm{CO}$. For $\rm{OH}$, which is commonly believed to be optically thin, we adopt a value of $4~\rm{cm^{-3}}$ for the critical excitation density (Goldsmith \& Li 2005).

For the disk-like and filamentary cores the path of integration and thus the column density also depends on the projection angle. For example, for a cylindrical core the path of integration will be larger when it is viewed with its axis of symmetry parallel to the line of sight rather than when it is seen with its axis of symmetry perpendicular to the line of sight. What is more, the molecular abundance distribution and the fact that we are only considering the portions of the core where $\rm{n_{tot}}\ge\rm{n_{eff}}$ when integrating, will have an effect on the resulting molecular column density map which also depends on the projection angle. Hence, we distinguish three cases; face-on, edge-on and intermediate angles. We have identified geometry-probing molecules that apply for each case {\em separately}. Consequently, the aspect ratio of the projected object as observed in dust emission determines the geometry-probing molecules that should be used.

\subsection{Face-on}\label{face_on}
Face-on, when all intrinsic 3D shapes appear circular, we only get one-dimensional information from the column density profiles for all intrinsic geometries (along the radius of the projected, circular object). The true 3D shape of a core has such a strong effect on the molecular column density profile that for a specific geometry the profiles appear similar for the majority of molecules or for group of molecules. In spherical cores, depletion is not obvious in column density maps in any of the molecules in our chemical network, not even for the ones that are most affected by it, such as $\rm{CO}$ and $\rm{HCO^{+}}$. The reason for this is that the path of integration shortens as the offset from the centre increases. Hence, lower abundances in the centre of spherical cores are counterbalanced by the larger integration paths. 

In disk-like cores, seen face-on, molecular column density profiles are usually flat since the path of integration is always the same and molecular abundances mostly change within an innermost, very thin disk. In cylindrical cores, we meet a variety of behaviours. For this intrinsic shape, column density profiles are either very centrally peaked or very centrally depleted. The difference with disk-like cores is that now abundances vary in a much larger portion of the core. If, for example, we consider a depletion-affected molecule, then in cylindrical cores we integrate through more gas where the abundance of that molecule is low than we do in disk-like cores.

Therefore, we need a way to measure the quantitative differences of molecular profiles. To do so, we define two parameters. The first parameter $\Delta$ is defined to be the ratio of mean values of molecular column density in two different regions of the core. These two regions are defined based on the {\em total column density} traced by dust emission maps. More specifically, the numerator is the mean molecular column density inside a contour that marks 50\% of the maximum total column density. The denominator is the mean of the molecular column density in the extent of the core where the total column density is higher than a background value, i.e.:

\begin{figure*}
\includegraphics[width=2.1\columnwidth, clip]{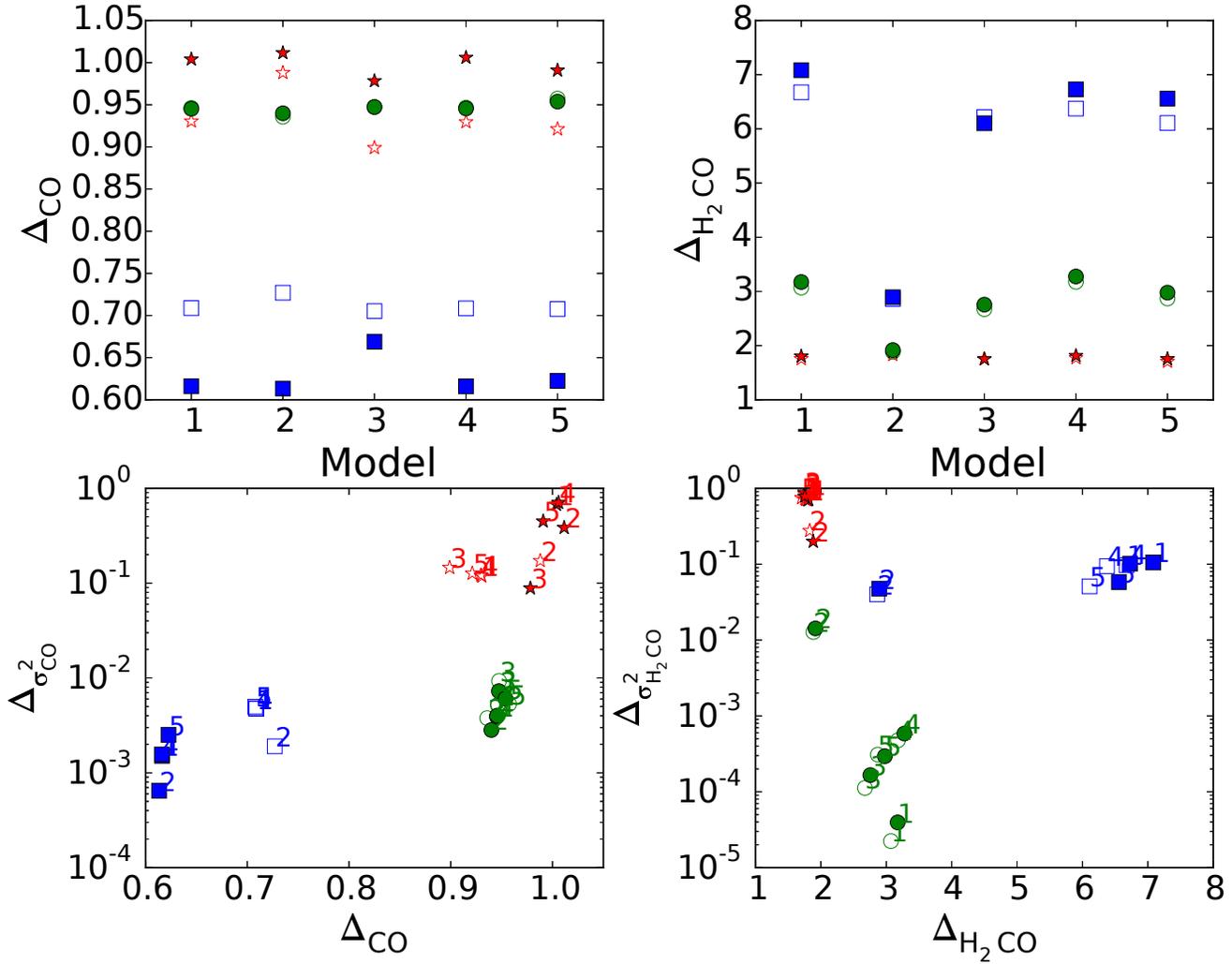}
\caption{Same as in Figure~\ref{main_combo2} for $\rm{CO}$ (left column) and $\rm{H_2CO}$ (right column). Numbers denote models with different parameters (see Table~\ref{Models}) and symbols represent different intrinsic geometries, as in Figure~\ref{main_combo2} (star=disk-like, circle=sphere, and square=cylindrical). We use open symbols to plot the results where we considered $\rm{n_{crit}}$ instead of $\rm{n_{eff}}$ when producing the molecular column density maps.} 
\label{multipanel_final2}
\end{figure*}

\begin{equation*}
\Delta=\frac{{\overline{N_X} ~inside ~a ~contour ~that \atop ~marks ~50\% ~of ~the ~maximum ~of ~N_{tot}}}{{\overline{N_X} ~in ~the ~extent ~of ~the \atop~core ~where ~N_{tot}~\ge ~background ~value}}
\end{equation*}
where X is a molecular species and the background total column density value is set to  $7\times10^{21}$ (K{\"o}nyves  et al. 2013)\footnote{We have verified that results do not change significantly if the contour in which we compute the numerator is changed to 25\% or 70\% of the maximum of the total column density and the background column density is changed a factor of two above and below the value adopted here.}. The definition of $\Delta$ is also shown schematically in Figure~\ref{Delta_explained3}. Here we show the total column density map (left panel) and the column density map of $\rm{CO}$ (right panel) for the spherical core. We have also overplotted the contour marking 50\% of the maximum of the total column density with a green circle and the one marking the region of the core with $\rm{N_{tot}} \ge \rm{7 \times 10^{21}}$ with a red circle in both maps.

The second parameter $\Delta_{\sigma^2}$ is the ratio of the variance of the molecular column density profiles in the same two regions of the core. Hence, while $\Delta$ quantifies whether a molecular column density profile is centrally peaked ($> 1$), depressed ($< 1$) or flat ($\approx 1$), $\Delta_{\sigma^2}$ quantifies how much a profile changes from the central to the outer regions of the core. Since the parameter $\Delta$ is a ratio of mean values, it can be confidently computed even from low resolution observations. However, to adequately describe the variation of molecular column density from the inner to the outer regions of the core and thus the parameter $\Delta_{\sigma^2}$, high angular resolution data are required.

%----------------------------------------------------------------------%
In the upper panel of Figure~\ref{main_combo2} we plot the parameter $\Delta$ for $\rm{OH}$ for each run and for each intrinsic geometry separately. We plot our results with red stars for the disk-like core, with blue squares for cylindrical core models, and with green circles for spherical cores. We present only the results obtained considering $\rm{n_{eff}}$ but results considering $\rm{n_{crit}}$ are identical since both values are very low. For our cylindrical core models, the column density profiles of $\rm{OH}$ are centrally depressed and clearly separate from the other two intrinsic shapes. The profiles of both the disk-like and spherical models are only slightly centrally depressed and the parameter $\Delta$ alone cannot break the degeneracy. However, when we plot the parameter $\Delta_{\sigma^2}$ against $\Delta$ in order to also include how the profiles change, the three geometries clearly separate. This is shown in the lower panel of Figure~\ref{main_combo2} where the numbers next to each symbol denote runs with different parameters. Two more molecules, $\rm{CO}$ and $\rm{H_2CO}$, that can be used in the same manner are shown in Figure~\ref{multipanel_final2}. In this Figure, we use open symbols to plot the results obtained with $\rm{n_{crit}}$. Thus, {\em the intrinsic geometry of a circular object can be identified by plotting the parameter $\Delta_{\sigma^2}$ against the parameter $\Delta$ for specific molecules. Different intrinsic 3D shapes will occupy different regions of the plot.}
%Complementary combinations may be used to first distinguish between spherical and cylindrical geometry and then determine if the cloud has a disk-like or a filamentary shape. Such combinations are ($CO$, $C_2D_2$), ($CHD$, $CH^{+}_3$), ($HCO$, $CH^{+}_3$), ($C_2HD$, $CO$), ($OH$, $C_2HD$), ($NHD_2$, $NH$), ($DCN$, $OH$), ($CO$, $H_2O$), ($HCO^{+}$, $C_2HD$) and are presented in .

\subsection{Edge-on}\label{edge_on}

%============================================================================
\begin{figure}
\centering
\includegraphics[width=1.0\columnwidth, clip]{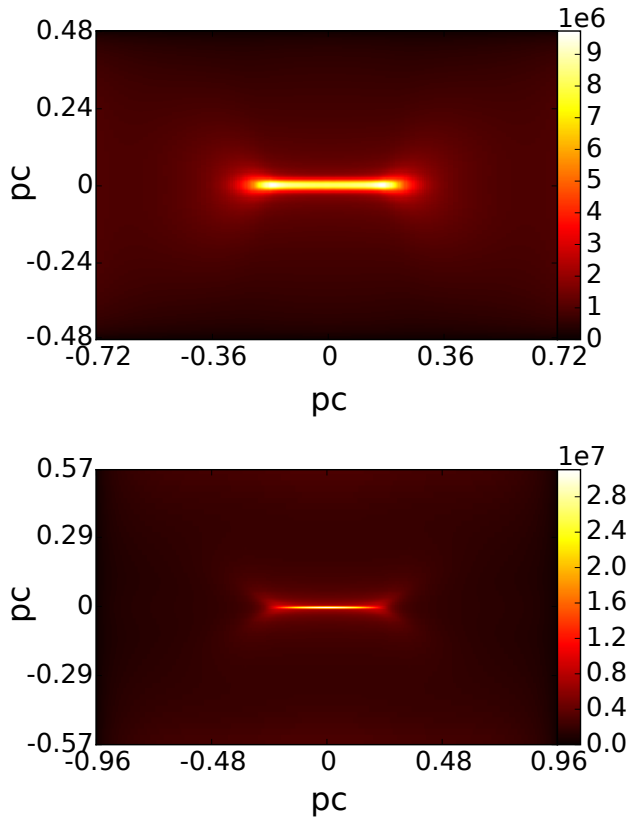}
\caption{$\rm{CD_2}$ column density maps for a cylindrical (upper panel) and an disk-like (lower panel) core. The maps closely resemble those of the total column density. \label{CD2_edge_on2}}
\end{figure}

\begin{figure}
\centering
\includegraphics[width=1.0\columnwidth, clip]{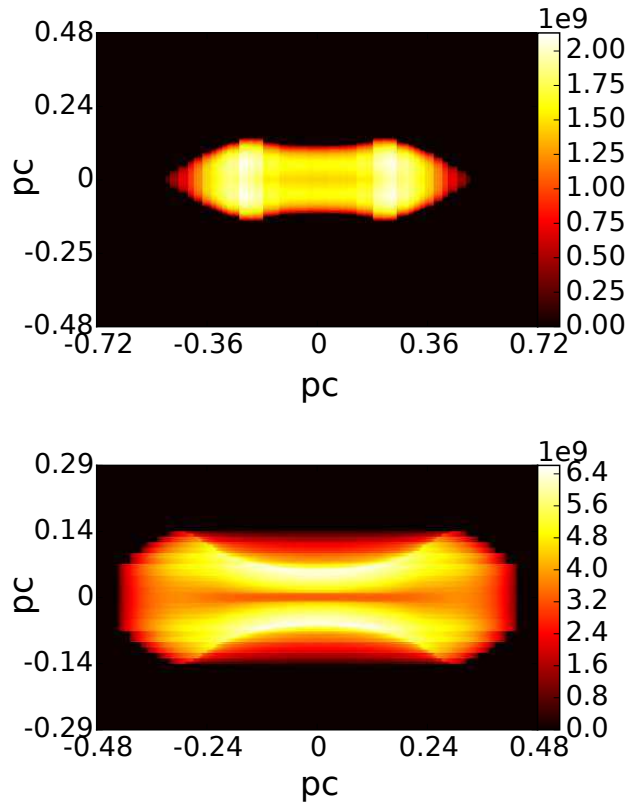}
\caption{Zoomed $\rm{N_2H^{+}}$ column density maps for a cylindrical (top) and a disk-like (bottom) core seen edge-on. Temperature and ionization rate are those of our reference run. The disk-like core would be observed as two parallel filaments whereas the cylindrical core would split into two horseshoe-shaped cores. 
\label{n2h_p_edge_on2}}
\end{figure}

When disk-like and cylindrical clouds are observed edge-on (i.e. for aspect ratios $\le$ 0.15), distinguishing them from spherical clouds is trivial. What is more, we now get two dimensional information (e.g. along the 2 principal axes of the projected object) and classification of clouds can be made without use of the parameter $\Delta$. Instead, a simple comparison of 2D emission maps from observations with column density maps from simulations for key molecules is sufficient. 

{\em All fully deuterated hydrocarbons consisting of one carbon atom trace the total column density very accurately} and thus the resulting maps for both a cylindrical and a disk-like core would be qualitatively indistinguishable from a dust continuum emission map. For $\rm{CD_2}$ this is shown in Figure~\ref{CD2_edge_on2}. Certain fully deuterated hydrocarbons consisting of two carbon atoms, such as $\rm{C_2D_2}$ also behave in the same manner. Partly deuterated hydrocarbons with one or more carbon atoms such as $\rm{C_2HD}$ and $\rm{CH_2D}$ may also be good density tracers but this is not a robust property against the parameter study. Since these molecules follow the total density we have not taken into account radiative transfer processes when producing their column density maps. If the critical densities of these molecules are much higher than $\rm{10^6 ~ cm^{-3}}$, i.e. the central density of all three modelled cores, then they will not be detected in a core of that evolutionary stage. However, if their critical densities are smaller than $\rm{10^6 ~ cm^{-3}}$ then they will continue to trace the central region and their molecular column density maps will qualitatively resemble those of the total column density in all projection angles and thus suffer from the same degeneracies with respect to the intrinsic shape.

Qualitative differences between cylindrical and disk-like clouds are observed in the column density maps of certain molecules with low critical densities. In $\rm{CH_3N}$ and $\rm{OH}$, disk-like cores manifest themselves as a depletion hole. In contrast, an irregular shape with no clear structures results when the core is cylindrical. Unfortunately, the differences are not large enough in order for these molecules to be used as probes of the intrinsic geometry.

\begin{figure*}
\centering
\includegraphics[width=1.9\columnwidth, clip]{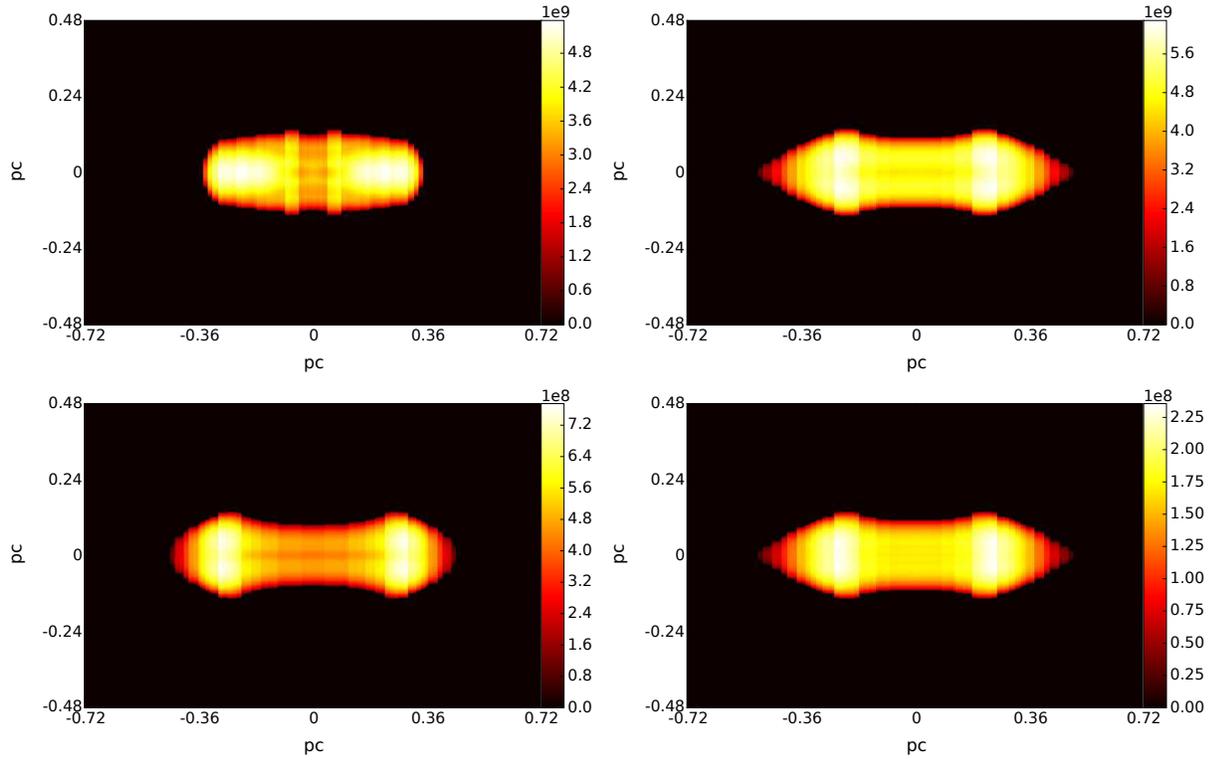}
\caption{$\rm{N_2H^{+}}$ column density maps for a {\em filamentary} core as seen edge-on for various sets of parameters. Top left is run 2, top right is run 3, bottom left is run 4 and bottom right is run 5. See Table~\ref{Models} for details on the parameters of each run.
\label{n2h_p_prolate_rest_params2}}
\end{figure*}

\begin{figure*}
\centering
\includegraphics[width=1.9\columnwidth, clip]{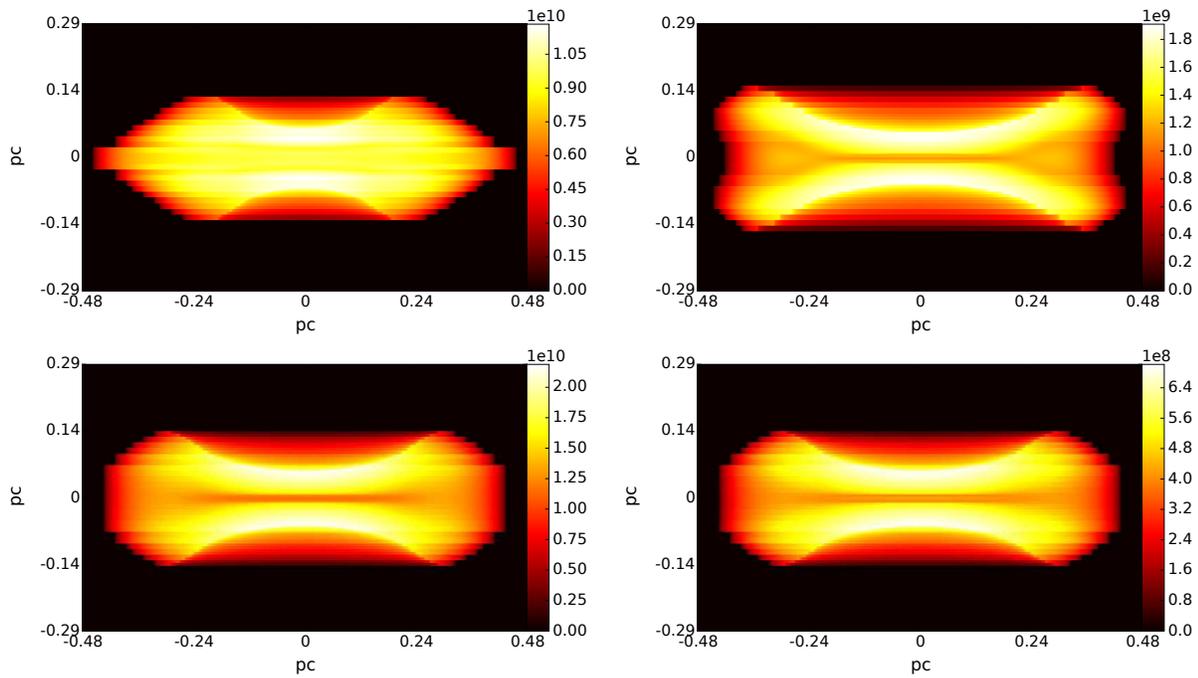}
\caption{Same as in Figure~\ref{n2h_p_prolate_rest_params2} but for our {\em disk-like} models.
\label{n2h_p_oblate_rest_params2}}
\end{figure*}

In Figure~\ref{n2h_p_edge_on2} we show zoomed $\rm{N_2H^{+}}$ column density maps of the cylindrical (upper panel) and disk-like (lower panel) cores for our reference run. In $\rm{N_2H^{+}}$ a disk-like core would be observed as two semi-parallel, slightly bent filaments. Cylindrical cores, which are a better approximation to a filamentary cloud, will either be seen as two separate elliptical objects or as continuous elongated structures, similar to but broader than in dust emission maps. Column density maps of $\rm{N_2H^{+}}$ for the rest of our parameter study runs are shown in Figures~\ref{n2h_p_prolate_rest_params2} \&  \ref{n2h_p_oblate_rest_params2}. When the $\rm{N_2H^{+}}$ column density maps of disk-like cores are produced by considering $\rm{n_{crit}}$ instead of $\rm{n_{eff}}$ this {\em splitting effect} is still prominent. 

The effect does not uniquely occur for $\rm{N_2H^{+}}$. Other molecules that follow the same behaviour are $\rm{NH_3}$ and $\rm{C_3H_2}$. The column density maps of these molecules are very similar to those of $\rm{N_2H^{+}}$ for both intrinsic shapes. For $\rm{NH_3}$ this splitting in molecular column density maps of disk-like models occurs independently of whether we include radiative transfer effects in our analysis. In contrast, for $\rm{C_3H_2}$ the effect is visible only when we consider the regions of the core with $\rm{n_{tot}}\ge\rm{n_{eff}}$. 

As a result, whether the {\em splitting effect} is visible in a molecular column density map of our disk-like core also depends on the critical density of the molecule under consideration. In contrast to $\rm{C_3H_2}$, the effect is visible in $\rm{CH_3CN}$ if we neglect the critical density when integrating to get its 2D projection map but it is not visible when we properly take it into account. In the latter case, the $\rm{CH_3CN}$ column density maps of both the disk-like and cylindrical core models resemble those of the total column density. If the critical density is neglected, the {\em splitting effect} is also seen in the molecular column density maps of $\rm{HCO^{+}_2}$, $\rm{H_2CO}$, $\rm{H_2DO^{+}}$, $\rm{CD^{+}_3}$, $\rm{C_3H_2D^{+}}$, $\rm{CHD^{+}_2}$, $\rm{CH_2D^{+}}$, $\rm{CH_4}$, $\rm{CO_2}$, and $\rm{DCO^{+}_2}$ provided of course that the shape of the core is that of the disk.  Thus, these molecules are promising candidates in which the {\em splitting effect} may be seen, especially if they have transitions with low effective densities.

\begin{figure}
\centering
\includegraphics[width=1.0\columnwidth, clip]{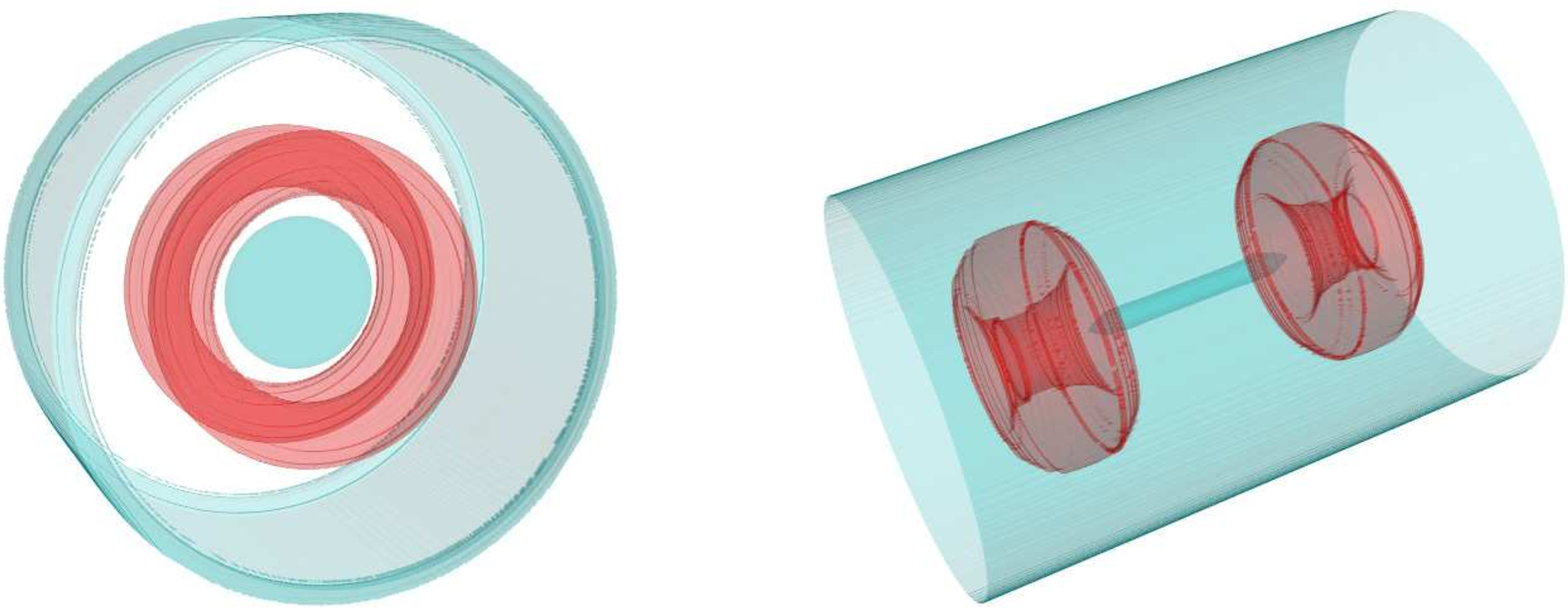}
\caption{$\rm{N_2H^{+}}$ 3D abundance plots for a disk-like (left) and a cylindrical (right) core. Isosurfaces are at 35\% (blue shaded isosurfaces) and 90\% (red shaded isosurfaces) of the maximum $\rm{N_2H^{+}}$ abundance. The inner, low abundance disk will have a larger radius than that of the inner, low abundance cylinder and as a result the effect of depletion will be more severe in the column density map of a disk when it is observed edge-on. 
\label{depletion_effect}}
\end{figure}

\begin{figure}
\centering
\includegraphics[width=1.0\columnwidth, clip]{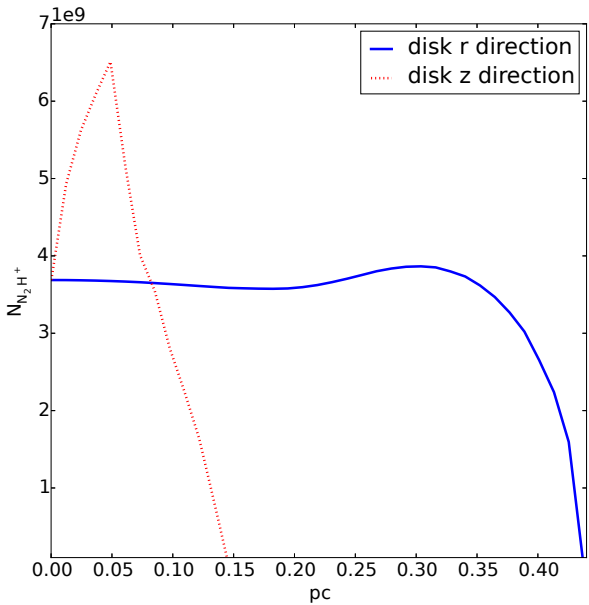}
\caption{$\rm{N_2H^{+}}$ column density profiles for a disk-like core. The gradient in the z direction from the peak to the centre of the disk-like core (dotted red line) is larger than that in the radial direction (solid blue line). As a result, a disk-like cores will not be observed as an empty ellipse when observed in $\rm{N_2H^{+}}$.  
\label{n2h_p_column_density_profiles3}}
\end{figure}

This {\em splitting effect} can be understood if depletion is considered. Despite the fact that $\rm{N_2H^{+}}$ is a high-density tracer (Tafalla et al. 2002), at very high densities depletion will eventually take over. Bergin et al.(2002) took $\rm{N_2H^{+}}$ (1 - 0) observations of the well studied prestellar core B68 using the IRAM 30 m telescope (beam size 25"). They found an $\rm{N_2H^{+}}$ depletion hole towards the centre of the core. Pagani et al. (2007a) \& Pagani et al. (2007b) observed the prestellar core L183 with the same telescope and found that $\rm{N_2H^{+}}$ was depleted by a factor of 6 in the inner regions of the core. Chitsazzadeh et al. (2014) observed the prestellar core L1689-SMM16 and also found a decrease in abundance of $\rm{NH_3}$ and $\rm{N_2H^{+}}$ towards higher densities. Di Francesco et al. (2004) found evidence of $\rm{N_2H^{+}}$ depletion towards the Ophiuchus A core. In agreement to these observations the abundance of $\rm{NH_3}$ and $\rm{N_2H^{+}}$ in our chemical models drops inside the high-density region. In Figure~\ref{depletion_effect} we show, in 3D, the $\rm{N_2H^{+}}$ abundance distribution for the disk-like (left) and cylindrical (right) core models. Isosurfaces are at 35\% (blue isosurfaces) and 90\% (red isosurfaces) of the maximum $\rm{N_2H^{+}}$ abundance. The inner blue shaded isosurface of the disk-like core has a larger radius than that of the cylindrical core. Thus, for a thin disk seen edge-on, the effect in molecular column density will be more severe than that for a cylindrical cloud, since the path of integration along a line of sight passing through the high-density region is larger. Furthermore, for a disk-like core the high-abundance region of $\rm{N_2H^{+}}$ will be parallel to the actual high-density region, whereas the opposite is true for a cylindrical cloud. Consequently, in such a map, the $\rm{N_2H^{+}}$ column density will probe the regions above and below the disk. With this reasoning, one might expect that the column density of $\rm{N_2H^{+}}$ would also probe the regions on either side of the high-density region of the disk. Hence, if the regions surrounding the disk were also connected, the resulting projected shape would be that of an empty ellipse. However, the gradient from the $\rm{N_2H^{+}}$ column density peak towards the centre of the cloud is sharper in the z direction (see Figure~\ref{n2h_p_column_density_profiles3}) and the net result is two parallel elongated structures. 

The {\em splitting effect} that occurs for disk-like cores might have already been observed. Fern{\'a}ndez-L{\'o}pez \etal (2014) observed the Serpens South molecular cloud in $\rm{N_2H^{+}}$ ($J=1\rightarrow 0$) using CARMA. They found that what appeared as a single filamentary structure in dust continuum was actually composed of two or three $\rm{N_2H^{+}}$ filaments. In a subsequent study of Serpens Main and with the same tracer, Lee et al. (2014) also found  two separate filaments inside what appeared to be a filamentary structure in dust emission. They argued that with CARMA's higher resolution they were able to resolve substructures of the filamentary objects observed with $Herschel$. 

\begin{figure}
\centering
\includegraphics[width=1.0\columnwidth, clip]{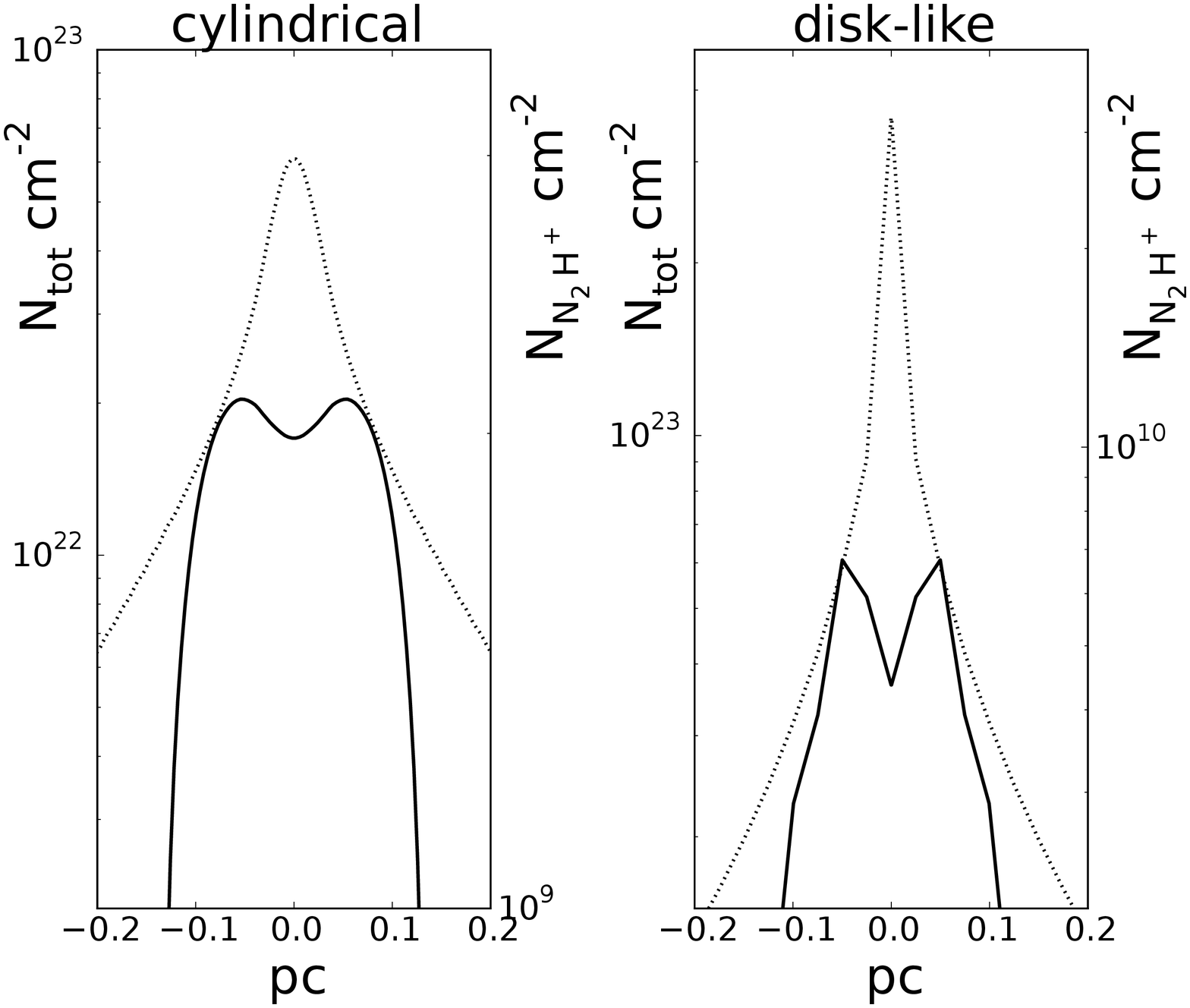}
\caption{Left: Total column density radial profile (dotted line) and $\rm{N_2H^{+}}$ column density radial profile (solid line) for a cylindrical core. Right: Total column density axial profile (dotted line) and $\rm{N_2H^{+}}$ column density axial profile (solid line) for a disk-like core. A scenario in which the $\rm{N_2H^{+}}$ filamentary structures have half the width from the filamentary structures observed in dust emission can more easily be realized if the intrinsic geometry of the core is disk-like rather than cylindrical. 
\label{profiles_cd_vs_N2H_p_cd}}
\end{figure}

Fern{\'a}ndez-L{\'o}pez \etal (2014) also estimated the width of the $\rm{N_2H^{+}}$ filaments. They found that these structures had approximately half the width of the filaments observed in dust continuum. Figure~\ref{profiles_cd_vs_N2H_p_cd} shows the total column density and $\rm{N_2H^{+}}$ column density profiles, parallel to the short axis, for our cylindrical (left) and disk-like (right) core models. If the true 3D shape of the structure observed by Fern{\'a}ndez-L{\'o}pez \etal (2014) were that of a cylinder then, due to the fact that no splitting occurs for such an intrinsic geometry, the $\rm{N_2H^{+}}$ profile would be broader, leading to a larger width, in contradiction to their results. If however, the intrinsic geometry of that object were disk-like, because of the splitting due to depletion, the two apparent $\rm{N_2H^{+}}$ filaments would have smaller widths (right panel of Figure~\ref{profiles_cd_vs_N2H_p_cd}). Sugitani et al.(2011) took polarization measurements at the region observed by Fern{\'a}ndez-L{\'o}pez \etal (2014). They found that the magnetic field was perpendicular to the filamentary structure where the splitting occurs.  

\begin{figure}
\includegraphics[width=1.0\columnwidth, clip]{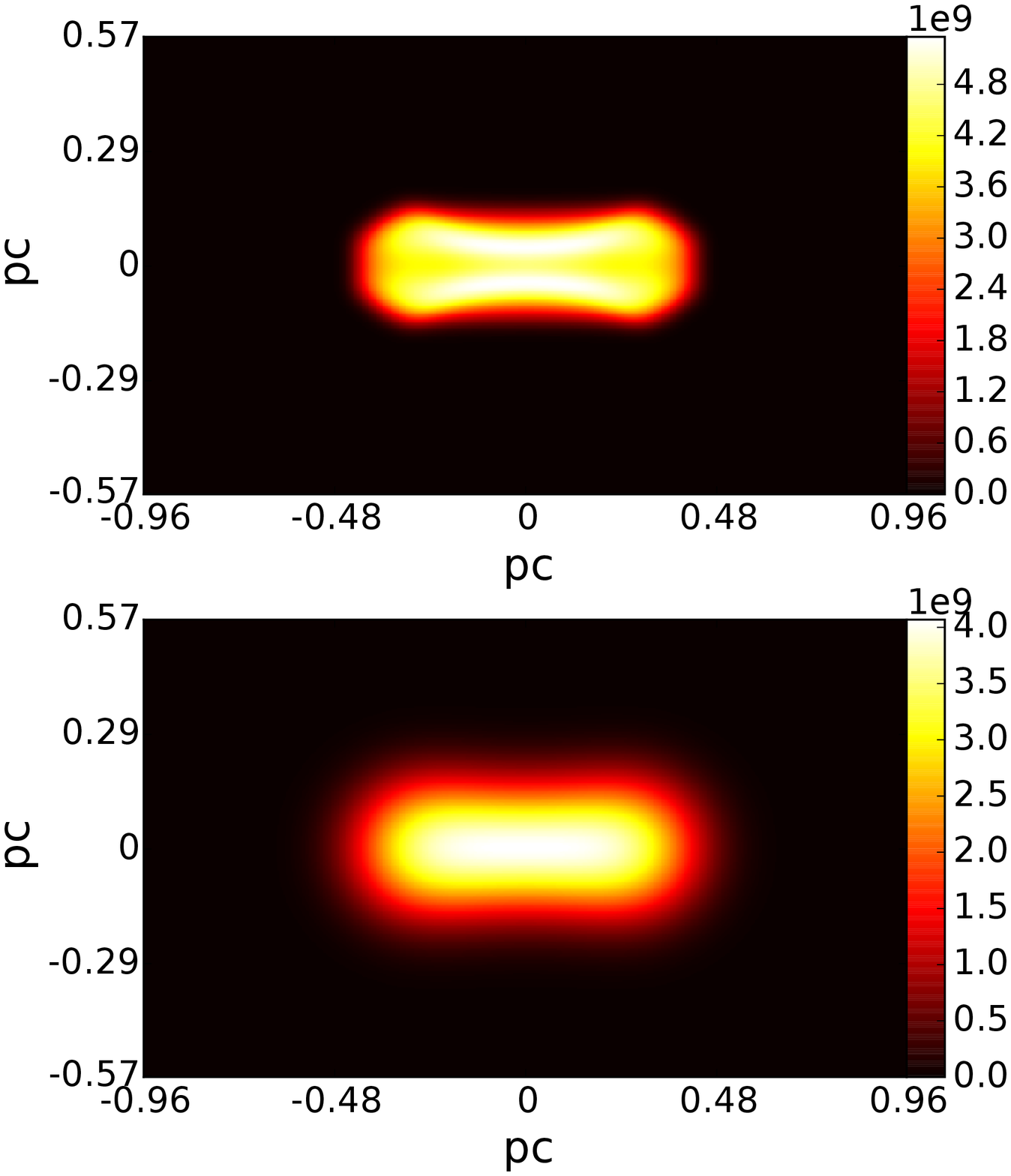}
\caption{$\rm{N_2H^{+}}$ column density maps for the disk-like core as observed edge-on convolved with two different beam sizes. In the upper panel the map was convolved with a beam size equal to that of CARMA and in the lower panel with $Herschel's$ 350$~\mu m$ beam size. At $Herschel's$ resolution the "splitting effect" is not visible and $\rm{N_2H^{+}}$ emission appears to peak at the z=0 plane.
\label{N2H_p_resolution}}
\end{figure}

As a proof of concept we have created the $\rm{N_2H^{+}}$ column density map of our disk-like core as seen edge-on at the highest resolution possible from the simulations. The core was then "placed" at a distance of 415 pc (i.e. the same distance adopted by Fern{\'a}ndez-L{\'o}pez et al. 2014 for the Serpens South Molecular Cloud) and the image was convolved using a Gaussian filter assuming a beam size equal to that of CARMA (7") (upper panel of Figure~\ref{N2H_p_resolution}). We then convolved the original map to $Herschel's$ 350$~\mu m$ beam size (25") (lower panel of Figure~\ref{N2H_p_resolution}). In agreement with the observational results of Fern{\'a}ndez-L{\'o}pez et al. (2014) the splitting is no longer visible when $\rm{N_2H^{+}}$ emission is convolved to $Herschel's$ resolution (see their Figure 3 where they perform the same analysis using observations).

\subsection{Intermediate projection angles}

Observationally, there is no way of determining the angle between the axis of symmetry of a core and the line of sight. Also, for the same projection angle, an ellipsoid resulting from a cylindrical cloud will have a different aspect ratio than an ellipsoid resulting from the projection of a disk-like core. Thus, in order to have a recipe that probes the true 3D shape of a core, it is only meaningful to relate a parameter to the aspect ratio of the projected shape. That parameter can again be the parameter $\Delta$ as defined in \S~\ref{face_on}. Here, there is no need to make use of the parameter $\Delta_{\sigma^2}$ since the degeneracy between disk-like and spherical cores is broken by dust emission maps alone.

In order to compute the aspect ratio of the projection of core, we first define the contour that marks 50\% of the maximum of the total column density. We then divide the lengths of the two principal axes determined by that contour. Thus, we define both the parameter $\Delta$ and the aspect ratio in a self-consistent manner.

Limited observational resolution can cause a projected object look rounder and can also change the apparent peakness of a molecular column density profile. Hence, for a molecule to be a good geometry tracer the value of the parameter $\Delta$ must be different for cylindrical and disk-like cores regardless of the resolution. To that end, results from simulations were convolved with a Gaussian filter assuming a beam size equal to 35" and a distance of 1 kpc. The value of the parameter $\Delta$ and the aspect ration were then computed at both instances. We have identified one molecule which satisfies these criteria. 

\begin{figure*}
\centering
\includegraphics[width=2.1\columnwidth, clip]{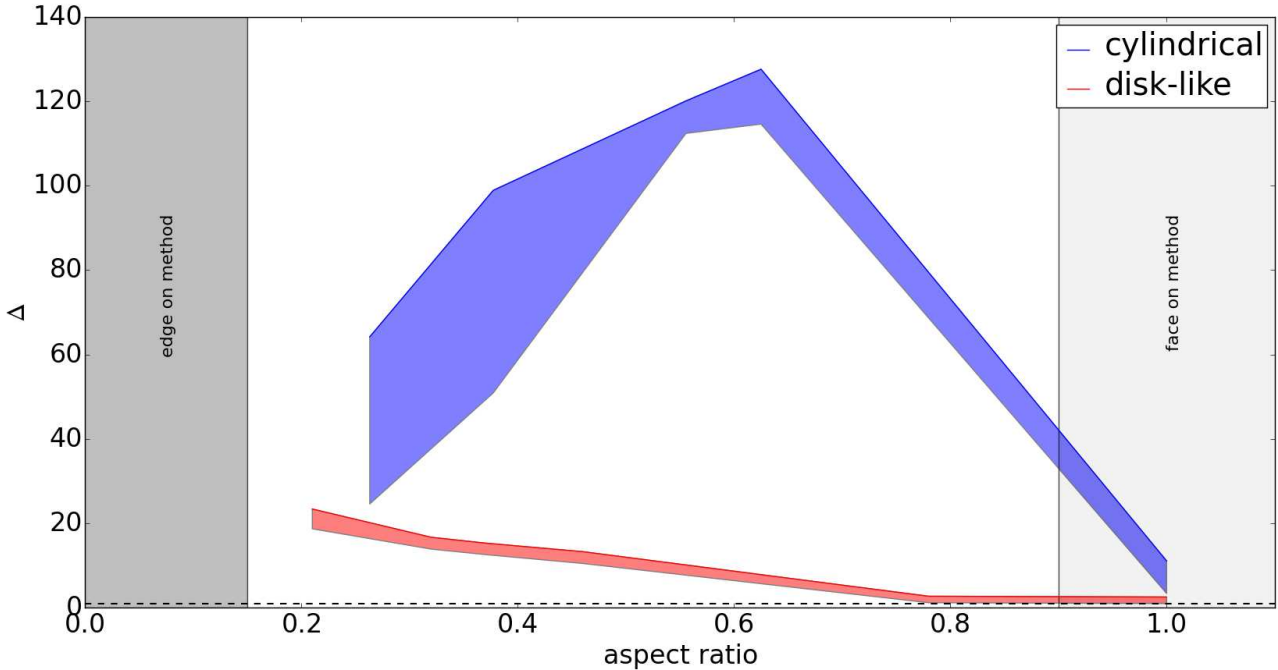}
\caption{The parameter $\Delta$ of CN as a function of the aspect ratio for disk-like and cylindrical models (red and blue shaded regions respectively) for different resolutions. The horizontal dotted line is at $\Delta=1$. Results are obtained considering $\rm{n_{eff}}$. For disk-like models $\Delta_{\rm{CN}}$ is close to unity, especially for aspect ratios $\ge$ 0.6 whereas for cylindrical models the value of the parameter $\Delta$ is much greater than 1 for all aspect ratios. \label{all_around_method_original2}}
\end{figure*}

In Figure~\ref{all_around_method_original2} we show the parameter $\Delta_{\rm{CN}}$ as a function of the aspect ratio. If the true shape of the core is that of a disk (red shaded region in Figure~\ref{all_around_method_original2}), $\rm{\Delta_{\rm{CN}}}$ is close to unity, especially for aspect ratios $\ge$ 0.6. In contrast, if the core is cylindrical (blue shaded region in Figure~\ref{all_around_method_original2}), the molecular column density profiles along both axes of the projected object are very centrally peaked and  $\Delta \gg 1$. This is the case for aspect ratios from 0.2 to unity, although the greatest differences in the values of the parameter $\Delta$ are found if the projected objects have aspect ratios $\approx$ 0.6. In Figure~\ref{all_around_method_original2} the upper boundaries of the shaded regions are obtained by considering the best possible resolution of the simulations. The value of $\Delta$ and the aspect ratio for different combinations of beam sizes and distances fall inside these shaded regions provided that the resolution is better than that obtained with a beam size of 35" and distance 1 kpc. The results presented here are obtained by considering the effective excitation density $\rm{n_{eff}}$ of $\rm{CN}$ although results obtained by considering the critical density $\rm{n_{crit}}$ of $\rm{CN}$ are in very good agreement. 

The differences between cylindrical and disk-like cores seen in Figure~\ref{all_around_method_original2} can be understood as follows. For cylindrical clouds, the molecular abundance of $\rm{CN}$ will be higher in a cylindrical shell just outside the innermost high-density thin cylinder. When such a core is seen at an offset from exactly edge-on, higher abundances will be integrated over a longer line of sight. As a result, the molecular column density will peak even more in the central region and the parameter $\Delta$ will increase. In contrast, for disk-like cores the path of integration gets shorter as the viewing angle changes from exactly edge-on to exactly face-on and therefore the value of $\Delta$ will be closer to unity.

\section{Discussion and Conclusions}\label{conclusions}

In the simulations presented here, the initial abundances of all chemical compounds were set to zero. In order to examine to what extent this approximation is unphysical, we have also considered a run in which the chemistry is left to evolve for 1 Myr at a uniform density of $\rm{10^3 ~cm^{-3}}$ before collapse. We have confirmed that the abundances at a central density of $\rm{10^6 cm^{-3}}$ do not change significantly. However, as it has been pointed out by previous studies (Aikawa et al. 2001), if the pre-collapse phase lasts longer and/or is at a higher density, abundances will deviate more from the models without a pre-collapse phase.

The molecular column density maps presented in this paper are produced only from the portions of the cores where $\rm{n_{tot}}\ge\rm{n_{crit}}$ and $\rm{n_{tot}}\ge\rm{n_{eff}}$, where we considered the transitions corresponding to the lowest possible values for $\rm{n_{crit}}$ and $\rm{n_{eff}}$. These are the inner regions of the cores, away from the boundaries of the computational region. Furthermore, when defining $\Delta$ and $\Delta_{\sigma^2}$, we have not included values of molecular column density that coincided with regions with total column density below a background value. We are thus confident that our results are not affected by the boundaries. 

The disk-like and cylindrical cores as well as the spherical core are idealized shapes of what can be found in nature. Triaxial cores appear to be a better fit to observations in statistical studies although axisymmetric shapes are not ruled out (Tassis 2007). The {\em splitting effect} might not be visible in the same manner if we have a triaxial oblate spheroid instead of a disk-like geometry. The abundance of $\rm{N_2H^{+}}$ will continue to probe the regions above and bellow the high-density region, but for an oblate spheroid the path of integration also gets smaller as we move along the short axis. Hence, for an oblate spheroid the higher abundance of $\rm{N_2H^{+}}$ around the high-density region might be counterbalanced by the shorter path of integration. However, a scenario in which an oblate spheroid observed edge-on will split in 3 parallel elongated structures when seen in $\rm{N_2H^{+}}$ can also be realized. In such a scenario, the middle structure will be seen due to the fact that $\rm{N_2H^{+}}$ also probes the regions in front of and behind the depleted region. A dip might then be caused due to lower abundance and smaller integration path, and finally two elongated structures above and bellow the middle one will be seen, simply due to higher $\rm{N_2H^{+}}$ abundance. In any case, the occurrence or not of the {\em splitting effect} for an oblate spheroid depends on its intrinsic aspect ratio and the interplay between $\rm{N_2H^{+}}$ abundance and integration path.

What is more, the {\em splitting effect} will not be visible for cores in earlier evolutionary stages. For a disk-like core with central density $\rm{\sim10^4 cm^{-3}}$ only the innermost regions of the core will be visible in $\rm{N_2H^{+}}$ since its effective density is of that order. Nonetheless, in case the {\em splitting effect} is observed in a manner similar to that presented in this paper, we can safely conclude that the intrinsic 3D shape of the core resembles a disk. Likewise, if a core appears filamentary in dust emission, the central density is of the order of $\rm{\sim10^6 cm^{-3}}$, and is seen as two separate elliptical objects or as a broad continuous structure in a $\rm{N_2H^{+}}$ column density map, we can conclude that its true shape is a cylinder. Finally, if the core is seen as a circularly symmetric object we can apply the face on method to determine the true shape of the core. We intend to return to the problem of triaxial cores with supplementary 3D simulations, including magnetic fields, and with a post-processing analysis of our result with a radiative transfer code in a future publication.

For projection angles such that our simulated cylindrical and disk-like core models are seen as ellipses or round objects the effect of depletion is not severe. As a result, both cores would appear centrally peaked in $\rm{N_2H^{+}}$ emission, with $\rm{\Delta > 1}$. In fact, at low resolution the $\rm{N_2H^{+}}$ column density maps of the two cores would be qualitatively identical to their respective total column density maps. This is in agreement with the early survey of $\rm{N_2H^{+}}$ emission from Lee et al. (2001) which had a large beam size (52"). 

Whether a molecule is a good geometry-tracer does not simply depend on which gas density better traces. It rather depends on its abundance distribution. The majority of molecules mentioned in \S~\ref{edge_on} have an abundance that peaks at a density of approximately a few times $\rm{10^4~cm^{-3}}$ to $\rm{10^5~cm^{-3}}$. Their abundance then starts to decline for higher densities. In contrast, the abundance profile of $\rm{CN}$ has a plateau at slightly higher densities, although it too depletes at densities of the order of $\rm{10^6~cm^{-3}}$. If our cylindrical and disk-like cores are seen edge-on then $\rm{N_2H^{+}}$ and $\rm{NH_3}$ are proven good geometry traces because they exhibit the "right amount" of depletion. If, for the same projection angle, the cores were observed in $\rm{CO}$ then they would both appear as depletion holes with no qualitative differences amongst them. 

For a random projection angle we derive the aspect ratio by dividing the major and the minor axis of the ellipse defined by the contour that marks 50\% of the maximum of the total column density. This is the same contour we use to define the parameter $\Delta$. We have confirmed that if we estimate the aspect ratio with a method based on the first and second moment of the flux density (e.g. Tassis et al. 2009) our results are not affected. However, for self-consistency, we recommend to define the aspect ratio and the parameter $\Delta$ in the same manner.

Our method can be summarized in the following steps:

\begin{itemize}
\item{From dust continuum observations determine the aspect ratio of the core by defining a contour that marks 50\% of the maximum of the total column density and dividing the two principal axes. The geometry-probing molecule that should be used is determined by the aspect ratio of the projected object.}
\item{For aspect ratios $\le$ 0.15 the geometry probing molecule will be $\rm{N_2H^{+}}$ (and/or $\rm{NH_3}$). 
\begin{enumerate}
\item{If the core appears as two parallel filamentary structures in the $\rm{N_2H^{+}}$ ($\rm{NH_3}$ ) column density map then its true shape is that of a disk.}
\item{If it is seen as a continuous, broader than in dust emission, structure then it is cylindrical-like.}
\end{enumerate}}
\item{When the aspect ratio of the core is in the range 0.15 $\sim$ 0.9 compute the parameter $\Delta$ for $\rm{CN}$ as defined in \S~\ref{face_on}.
\begin{enumerate}
\item{If the value of the parameter $\Delta$ is $\gg 1$ then the core is cylindrical-like.}
\item{If $\Delta$ is close to unity then the geometry of the core is disk-like.}
\end{enumerate}}
\item{For well resolved, centrally peaked, circular objects with polar symmetry that can be face-on projections of disk-like or filamentary cores or projections of spherical cores calculate the parameters $\Delta_{\sigma^2}$ and $\Delta$ for $\rm{OH}$ (and/or $\rm{CO}$, $\rm{H_2CO}$) as defined in \S~\ref{face_on}.
\begin{enumerate}
\item{When both $\Delta_{\sigma^2}$ and $\Delta$ are close to unity the intrinsic shape of the core is disk-like.}
\item{If $\Delta$ is close to unity but $\Delta_{\sigma^2}$ $\ll$ 1 then the core is spherical.}
\item{If both $\Delta$ and $\Delta_{\sigma^2}$ $\ll$ 1 then the true shape of the core is cylindrical-like.}
\end{enumerate}}
\end{itemize}

\section*{Acknowledgements}

We thank G.V. Panopoulou, T. Mouschovias, V. Pavlidou, P. Goldsmith and N. Kylafis for useful suggestions and discussions. We also thank the anonymous referee for useful comments that helped improve this paper. The software used in this work was in part developed by the DOE NNSA-ASC OASCR Flash Center at the University of Chicago. 3D plots were created using Mayavi2 (Ramachandran \& Varoquaux 2012). For post processing our results we partly used yt analysis toolkit (Turk et al. 2011). K.T. and A.T. acknowledge support by FP7 through Marie Curie Career Integration Grant PCIG- GA-2011-293531 ``SFOnset".
A.T. and K.T. would like to acknowledge partial support from the EU FP7 Grant PIRSES-GA-2012-31578 ``EuroCal". K.W.'s work was carried out at the Jet Propulsion Laboratory, California Institute of Technology, under a contract with the National Aeronautics and Space Administration. K.W. acknowledges support from the NASA Origins of Solar System Program. Usage of the Metropolis HPC Facility at the CCQCN of the University of Crete, supported  by  the European Union Seventh Framework Programme (FP7-REGPOT-2012-2013-1) under grant agreement no. 316165, is also acknowledged.

\end{document}